\documentstyle[12pt,aaspp4, epsf]{article}
\def\gapprox{\mathrel{\vcenter{\offinterlineskip \hbox{$>$}
    \kern 0.3ex \hbox{$\sim$}}}}
\def\lapprox{\mathrel{\vcenter{\offinterlineskip \hbox{$<$}
    \kern 0.3ex \hbox{$\sim$}}}}
\def\refindent{\par\penalty-100\noindent\parskip=4pt plus1pt
               \hangindent=3pc\hangafter=1\null}

\slugcomment{Submitted to Ap.J.}
\begin{document}
\setlength{\baselineskip}{12pt}

\title{Global Magnetohydrodynamical Simulations of Accretion Tori}
\author{John F. Hawley}
\affil{Virginia Institute of Theoretical Astronomy, \\
Department of Astronomy, University of Virginia, \\
Charlottesville, VA 22903; jh8h@virginia.edu}

%\received{10/30/98}
\begin{abstract}

Global time-dependent simulations provide a means to investigate
time-dependent dynamic evolution in accretion disks.  This paper seeks
to extend previous local simulations by beginning a systematic effort
to develop fully global three-dimensional simulations.  The nonlinear
development of the magnetorotational instability is investigated using
a time-explicit finite difference code written in cylindrical
coordinates.  The equations of ideal magnetohydrodynamics are solved
with the assumption of an adiabatic equation of state.  Both a
Newtonian, and a pseudo-Newtonian potential are used.  Two
simplifications are also explored:  a cylindrical gravitational
potential (the ``cylindrical disk''), and axisymmetry.  The results
from those simulations are compared with fully three dimensional global
simulations.

The global simulations begin with equilibrium pressure supported
accretion tori.  Two different initial field geometries are
investigated:  poloidal fields that are constant along initial
equidensity surfaces, and toroidal fields with a constant ratio of gas
to magnetic pressure.  In both cases the magnetorotational instability
rapidly develops, and the torus becomes turbulent.  The resulting
turbulence transports angular momentum, and the torus develops an
angular momentum distribution that is near Keplerian.  A comparison
with axisymmetric simulations shows that in three dimensions the
magnetorotational instability can act as a dynamo and regenerate
poloidal field thereby sustaining the turbulence.  As previously
observed in local simulations, the stress is dominated by the Maxwell
component.  The total stress in the interior of the disk is $\approx
0.1$-0.2 the thermal pressure.  At late time the disks are
characterized by relatively thick configurations, with rapid
time-dependence, and tightly-wrapped, low-$m$ spiral structures.

\end{abstract}
\keywords{accretion, accretion disks - instabilities - magnetohydrodynamics}

\section{Introduction}

Accretion powers a wide range of energetic objects, systems ranging in
size from low-mass X-ray binary systems to the disks surrounding
supermassive black holes in active galaxies and quasars.  Theoretical
models of these systems must be, necessarily, highly simplified.  One
such simplification that is often employed is to assume a time
stationary disk.  But the rapid time-variability seen in space- and
ground-based observations emphasizes that accretion disks are highly
dynamic systems.  These observations highlight the need to solve the
fully time-dependent equations describing accretion disks, and that, in
turn, requires numerical techniques.

Recently, time-dependent numerical simulations have been applied to the
problem of accretion disk transport.  The discovery of the
magnetorotational instability (MRI) in accretion disks (Balbus \&
Hawley 1991) has elucidated the physical basis for this transport:
magnetohydrodynamic (MHD) turbulence.   At the same time, the
development of practical three-dimensional MHD codes brought
the study of disk MHD turbulence, transport, and evolution into
the computational domain.

To date, most of the numerical simulations of the MRI have been carried
out in a local disk approximation known as the shearing box (Hawley,
Gammie \& Balbus 1995, hereafter HGB).  In the shearing box model, one
considers a frame corotating with the angular velocity at a fiducial
radius $R_\circ$.  By restricting the computational domain to small
excursions from this fiducial radius, one can reduce the geometry (but
not the dynamics) to a simple Cartesian system while retaining tidal
and Coriolis forces.  Local shearing box simulations (e.g., HGB;
Hawley, Gammie, \& Balbus 1996, hereafter HGB2; Brandenburg et al.
1995; Matsumoto \& Tajima 1995; Stone et al. 1996) have demonstrated
that MHD turbulence is the nonlinear outcome of the MRI, and that
outward angular momentum transport is its natural consequence.  The
total stress $T_{R\phi}$ is dominated by the Maxwell, or magnetic,
component rather than the kinematic, or Reynolds stress.  In disks the
amplitude of this stress is often parameterized with the
Shakura-Sunyaev (1973) alpha formulation, $T_{R\phi}=\alpha P$, where
$P$ is the pressure.  In the local simulations typical stress values
range from $\alpha \approx 0.1$ to $0.01$.  These simulations are more
fully reviewed in Balbus \& Hawley (1998).

An important step in increasing the realism of numerical disk
studies is to move from this local approximation to fully global
disk simulations.  The difficulties of three dimensional global
simulations and, in particular, the demands such simulations place on
computer hardware, make this step challenging.  In general, the time-
and length-scale differences between the extended radial scales of an
accretion disk and the scale of the MHD turbulence are too great to be
fully resolved.  Nevertheless, global simulations are now possible
for somewhat restricted ranges in radius, height, and angle. 

Several global MHD disk simulations have already been done.  One
of the first, by Matusmoto \& Shibata (1997), followed the evolution of a
thick torus embedded in a weak vertical magnetic field.  What transpires
is highly dynamic:  the outer layers of the torus slough
inward in a process referred to as avalanche accretion.  This avalanche
accretion is the global consequence of the radial streaming motion
(the ``channel solution'') which is the nonlinear manifestation of the
vertical field instability in the local limit (Hawley \& Balbus 1992;
Goodman \& Xu 1994).  The vertical field winds up as it is twisted by
the torus rotation, and material is accelerated upward and outward
along the field lines.  

In subsequent work, Matsumoto (1999) evolved constant angular
momentum tori containing an initially toroidal field.  The
nonaxisymmetric MRI rapidly leads to turbulence.  Magnetic energy is
amplified and approaches a value of $\beta \equiv P/(B^2/8\pi) \approx
10$.  Again the onset and appearance of the nonlinear stage of the
instability are understandable in light of what has been seen in local
simulations.  In this simulation, however, the MRI has a global
consequence:  the resulting MHD stress transports angular momentum
outward, rapidly changing the angular momentum distribution from
constant with radius to nearly Keplerian.

Recently, Armitage (1998) performed a global calculation of a
cylindrical (no vertical structure) Keplerian disk using the ZEUS code
(Stone \& Norman 1992a,b).  The initial magnetic field was $B_z \propto
\sin(R)/R$, with the sine function chosen to vary once over the central
part of the disk.  This initial state is unstable, and turbulence
rapidly develops along with significant angular momentum transport;
$\alpha$ values are on order 0.1.  As in local simulations, the Maxwell
stress is several times the Reynolds stress.  The magnetic field in the
turbulent disk is dominated by the toroidal component, and the
turbulence itself is characterized by structures with small azimuthal
wavenumbers $m$.  Hawley \& Balbus (1999) similarly computed
cylindrical models of Keplerian disks and constant angular momentum
tori.  They found in all cases that the MRI rapidly developed leading
directly to turbulence and significant angular momentum transport.

Global simulations such as these have become possible only recently with
advances in available computer speed and memory.  As computer capacities 
continue to increase, ever more ambitious simulations will become possible.
The aim of this paper is to
begin a systematic effort to develop such three-dimensional global MHD disk
simulations.  To a large extent, this paper will focus on the technical
aspects of such simulations, and the gross dynamics of the MRI in
global disks.  What type of simulations are currently practical?  What
sort of diagnostics are useful?  What are the advantages and
limitations of simplifications such as axisymmetry or the cylindrical
approximation?  As these questions are addressed, however, several
significant conclusions about accretion torus dynamics will be
obtained.

The plan of this paper is as follows:  \S2 discusses technical aspects
of the global simulations, including the equations, the numerical
algorithm, test problems, initial conditions, and diagnostic
procedures.  In \S3 the results from a range of simulations are
presented.  Two important simplifications are investigated:
two-dimensional axisymmetry, and a cylindrical gravitational
potential.  These approximations reduce the computational complexity of
global simulations, and a comparison with the full three-dimensional
treatment allows their limitations to be assessed.  Simulations of
fully global three dimensional tori follow.  Finally \S4 summarizes the
conclusions from this work.

\section{Global Simulations}

\subsection{Equations and Algorithms}

The dynamical evolution of an accretion disk is governed by 
the equations of ideal MHD, i.e.,
\begin{equation}\label{mass}
{\partial\rho\over \partial t} + \nabla\cdot (\rho {\bf v}) =  0
\end{equation}
\begin{equation}\label{mom}
\rho {\partial{\bf v} \over \partial t} 
+ (\rho {\bf v}\cdot\nabla){\bf v} = -\nabla\left(
P + {B^2\over 8 \pi} \right)-\rho \nabla \Phi + 
\left( {{\bf B}\over 4\pi}\cdot \nabla\right){\bf B} 
\end{equation}
\begin{equation}\label{ene}
{\partial\rho\epsilon\over \partial t} + \nabla\cdot (\rho\epsilon {\bf v}) 
= -P \nabla \cdot {\bf v}
\end{equation}
\begin{equation} \label{ind}
{\partial{\bf B}\over \partial t} = 
\nabla\times\left( {\bf v} \times {\bf B} \right)
\end{equation}
where $\rho$ is the mass density, $\epsilon$ is the specific internal
energy, ${\bf v}$ is the fluid velocity, $P$ is the pressure, $\Phi$ is
the gravitational potential, and ${\bf B}$ is the magnetic field
vector.  In the present study we adopt an adiabatic equation of state,
$P=\rho\epsilon(\Gamma -1) = K\rho^\Gamma$, and ignore radiation
transport and losses.  Since there is no explicit resistivity or
physical viscosity, the gas can heat only through adiabatic
compression, or in shocks through explicit artificial viscosity, $Q$.  
Artificial viscosity appears as an additional pressure term in equation
(\ref{ene}), i.e., $-(P+Q)\nabla \cdot {\bf v}$ (Stone \& Norman 1992a).
The absence of physical shear viscosity in these equations is worth
noting.  This is the actual state of the plasma:  the molecular
viscosity is utterly negligible.  Accretion disks evolve due to
transport resulting from what is normally referred to as anomalous viscosity.
But this is not an additional term in the equations; transport results
from the dynamics contained within the above equations.

In some simulations below,  the code employs the usual Newtonian
gravitational potential, $\Phi = -GM/r$, where $r$ is the spherical
radius.  To deal with the inner boundary in a more realistic manner,
however, the code also makes use of the pseudo-Newtonian potential
(Paczy\'nsky \& Wiita 1980) to model the relativistic effects
associated with a Schwarzschild metric (e.g., minimum stable orbit).
This inner boundary permits supersonic (and super-Alfv\'enic) accretion
off of the inner radial grid, reducing the likelihood of unphysical
influences from the boundary conditions.  The pseudo-Newtonian potential
has the form
\begin{equation}\label{pwp}
\Phi = - {G M \over r-r_g}
\end{equation}
where $r_g$ is the ``gravitational radius'' (akin to the black hole
horizon), here set equal to 1.  For this potential,
the Keplerian specific angular momentum is 
\begin{equation}\label{pwl}
\ell_{kep} = (GMr)^{1/2} {r \over r-r_g} ,  
\end{equation}
and $\Omega R^2 = \ell$.  Here we set $r_g = 1$, and, 
for both gravitational
potentials, $GM=1$.  This determines the units of time in the simulation with
$\Omega =1$ at $R=1$ for a Newtonian gravitational potential.  All
times given will be reported in these units; the corresponding orbital
periods at locations in the tori will also be given where appropriate.

Equations (\ref{mass})-(\ref{ind}) are solved using time-explicit
Eulerian finite differencing.  The global disk code is written in in
cylindrical coordinates, $(R,\phi,z)$.  The center of the coordinate
system is excised, i.e., the radial coordinate begins at a nonzero
minimum $R_{\rm min}$; this avoids the coordinate singularities
associated with the axis.  In some simulations, the gravitational
potential is assumed to be cylindrical; this removes the vertical 
component of gravity,
and reduces the number of $z$ grid zones required.  When the vertical
component of gravity is included, the $z$ coordinate is centered on the
equatorial plane.  The angle $\phi$ is periodic and covers the full
$2\pi$ or some integer fraction of $2\pi$.  In some simulations, only
the two-dimensional axisymmetric system $(R,z)$ will be computed.
A schematic of the computational domain is shown in Figure 1.

Most of the three dimensional global simulations are run on parallel
computer systems with a version of the global code that uses the
Message Passing Interface (MPI) library for interprocess
communication.  The full computational domain is divided into
overlapping subgrids, one for each processor.  Boundary zone values are
passed between processors using explicit MPI routines.  In the
description of what follows, however, this subdivision will be ignored;
the physical domain size and the number of grid zones discussed will be
the totals.

The MHD algorithm now has a long history of use.  The magnetic field is
evolved with the constrained transport (CT) approach of Evans and
Hawley (1988) that preserves the constraint $\nabla \cdot B =0$.  The
CT scheme was designed to work in any general curvilinear coordinate
system, and here the addition of appropriate geometric terms easily
adapts the procedure to cylindrical coordinates.  The algorithm uses
information propagated along Alfv\'en characteristics to solve a
restricted set of characteristic equations for time-advanced fields and
electromotive forces within the CT framework.  This approach is known
as the Method of Characteristics Constrained Transport (MOCCT)
algorithm, and its first implementation is described in detail in Stone
\& Norman (1992b); the current version is described in Hawley \& Stone
(1995).

Considerable effort has gone into determining the most satisfactory
scheme for the individual numerical terms in the present application.
Local conservation of angular momentum is particularly important, and a
number tests using hydrodynamic Keplerian disks on a variety of grids
suggest that some numerical approaches are better than others.
Specifically, it is best to evolve $\rho \ell$, where $\ell$ is the
specific angular momentum, as a fundamental variable (as opposed to
angular velocity, or specific angular momentum alone) and to use
consistent advection (Norman, Wilson \& Barton 1980) where the
advection of all variables is tied to the advection of the density.  To
minimize errors associated with the coordinate singularity located at
$R=0$ we make use of a regularized form of the operator
$R^{-1}\partial_R$ (Evans 1986).

A variety of test simulations were used to verify code performance.
The hydrodynamic implementation was tested with Keplerian disks and a
series of simulations of constant angular momentum tori similar to
those studied by Hawley (1991).  Constant and near-constant angular
momentum tori are subject to particularly vigorous growth of the
Papaloizou-Pringle instability (Papaloizou \& Pringle 1984), especially
when such tori are ``slender,'' meaning that their cross section is
small compared to the radius of the pressure maximum.  Slender tori are
unstable to the principal mode of the Papaloizou-Pringle instability,
with growth rates near the maximum value.  These tests gave results
that were consistent with past simulations.

For an MHD test, the grid was moved out to large radius and an isolated
vertical magnetic field flux tube was embedded in a Keplerian flow.
This test is a three-dimensional version of the axisymmetric
simulations described in Hawley \& Balbus (1991).  Again the results
were fully consistent with the earlier work.

\subsection{Accretion Tori}

Most of the simulations discussed below consider the problem of the
nonlinear evolution of the magnetorotational instability in accretion
tori.  Tori have significant internal pressure gradients that balance
the gravitational and centrifugal accelerations in hydrostatic
equilibrium.  This pressure gives the torus a vertical thickness $H$
that can be comparable to its radius $R$.  Significant departures from
Keplerian angular momentum distributions are possible as the torus
becomes ever thicker; the limit is the constant angular momentum
torus.  A particularly useful feature of these tori for numerical
simulations is that they are well-defined equilibria that can be
completely contained on the finite domain of a computational
simulation.  This allows their evolution to proceed (at least
initially) independent of the grid boundary conditions.

For an adiabatic equation of state, a gravitational potential $\Phi$,
and an assumed rotation law $\Omega \propto R^{-q}$, the density in the
torus is determined by
\begin{equation}\label{torus}
{\Gamma K P\over (\Gamma - 1)\rho} = 
C -\Phi - {1\over (2q-2)} {\ell_k^2\over R^{2q-2} }
\end{equation}
where $\ell_k$ is the Keplerian angular momentum at the pressure
maximum and $C$ is a constant of integration that establishes the zero
pressure surface of the torus.  A given torus is specified by the
angular velocity gradient $q$ (with $q=2$ corresponding to constant
angular momentum), the radial location of the pressure maximum
$R_{kep}$, and the radial location of the inner edge of the torus,
$R_{in}$ (which determines $C$).  Outside of the torus the grid is
filled with a cold gas with low density.  A floor value of the density
and internal energy is applied to ensure that these zones do not become
evacuated, or develop extremely high sound speeds.  The floor value of
density should be set low enough so that its contribution to the total
mass or energy is not significant at any place on the grid.  Here the floor
value is $10^{-3}$ and the typical torus initial density maximum is
$\sim 10$.

Tori are known to be subject to the purely hydrodynamic global
instability discovered by Papaloizou \& Pringle (1984).  The
Papaloizou-Pringle instability has been extensively investigated both
with perturbation analyses, and with several numerical simulations.
The mechanism of the instability was elucidated by Goldreich, Goodman
\& Narayan (1986) and involves the establishment of a global wave that
gains energy through the exchange of angular momentum between the inner
and outer region of the disk.  Amplification requires wave reflection
at at least one disk boundary.  The nonlinear evolution of the
Papaloizou-Pringle instability has been followed for both slender
(Hawley 1987), and relatively wide (Blaes \& Hawley 1988; Hawley 1991)
tori.  In the slender torus case, the instability saturates as a highly
nonaxisymmetric orbiting fluid ellipse (the ``planet'' solution; Hawley
1987) which itself proves to be an equilibrium solution (Goodman,
Narayan, Goldreich 1987).  Wider tori saturate in spiral pressure waves
(Blaes \& Hawley 1988; Hawley 1991).

Torus linear stability and, as we shall see, nonlinear dynamics are
quite different when a magnetic field is present.  The largest growth
rate of the MRI greatly exceeds those typical of the hydrodynamic
Papaloizou-Pringle instability.  Linear stability analyses bear this
out.  Curry \& Pudritz (1996) and Ogilvie \& Pringle (1996) have done
global linear MHD stability analyses of cylindrical systems ($1/R$
gravitational potential) with a variety of initial angular momentum
distributions.  They find rapidly growing unstable modes for a wide
range of equilibria and field strengths.

Although global linear analyses are most appropriate for strong
magnetic fields whose most unstable wavelengths are comparable to the
size of the torus, weak fields have unstable wavelengths that are much
smaller than the torus dimensions.  The essential physics of the MRI in
these circumstances is local, and the nonlinear evolution should be
similar to that seen in the local shearing box simulations.  This is,
of course, a testable proposition.  In any case, global simulations
will provide information regarding the ultimate consequences of the MRI
for the evolution, and, indeed, existence of thick tori.  Obviously, the
initial equilibrium torus cannot persist in the face of vigorous
instability and the resulting angular momentum transport.  Its overall
global evolution, however, may present novel features, and the end
state of its evolution may represent a more realistic accretion disk
structure.

\subsection{Diagnostics}

A global simulation will involve a grid with a million to a billion
zones, run for possibly hundreds of thousand of timesteps.  A major task is
to develop useful diagnostics that  concisely and adequately
characterize the essential behavior of an evolving disk in such a simulation.

The simplest type of diagnostic is the volume integral of a 
quantity over the entire grid.  For example, the total mass is
\begin{equation}\label{totmass}
M = \int \rho R dR d\phi dz.
\end{equation}
Time histories are computed of this quantity and 
others such as total kinetic and
magnetic energies (by component), and the radial mass flux
\begin{equation}\label{mdot}
\langle \dot M\rangle = \int \rho v_r R d\phi dz
\end{equation}
through both the inner and outer boundaries.

Standard steady-state thin disk models are often expressed in terms of
vertically averaged quantities.  Here such values can be
computed by integrating
density-weighted quantities over $z$ and $\phi$.  The 
vertically- and azimuthally-averaged mass density is
\begin{equation}\label{sigma}
\Sigma (R) = {\int \rho R d\phi dz \over \int R d\phi dz}.
\end{equation}
The averaged mass flux, $\langle \dot M \rangle $, is defined by
(\ref{mdot}),
and one can construct
similar averages for other quantities.  For non-Keplerian disks the
radial dependence of the average specific angular momentum 
will be of interest, and this can be computed from
\begin{equation}\label{el}
\langle \ell \rangle= \int \rho \ell R d\phi dz /\Sigma .
\end{equation}
The stress $\langle T_{R\phi}\rangle$ consists of a magnetic component
(the Maxwell stress)
\begin{equation}\label{maxwell}
\langle T_{R\phi}^{Max}\rangle = {\int (B_R B_\phi / 4\pi) R d\phi dz 
\over \int Rd\phi dz} ,
\end{equation}
and a kinematic component (the Reynolds stress)
\begin{equation}\label{Rey}
\langle T_{R\phi}^{Rey}\rangle = 
{\int \rho v_R \delta v_\phi Rd\phi dz\over \int Rd\phi dz}.
\end{equation}
The value of the perturbed angular velocity $\delta v_\phi$ is
defined for a Keplerian disk by
$\delta v_\phi = v_\phi - R\Omega_{Kep}$.  For other
angular momentum distributions, including average angular momentum
distributions that change with time, we adopt an alternative definition for
the perturbed orbital velocity $\delta v_\phi$
in terms of the difference between the total instantaneous angular
momentum flux, and the mass flux times the average angular momentum.
The difference then represents the excess or deficit angular momentum
transport due to orbital velocity fluctuations compared to the mean.
Specifically,
\begin{equation}\label{rstress}
\langle \rho v_R \delta v_\phi \rangle =
\langle \rho v_R v_\phi \rangle
- \langle \rho v_R \rangle \langle \ell \rangle / R .
\end{equation}

Together the Maxwell and Reynolds stress make up the total stress,
which, in the Shakura-Sunyaev (1973) parameterization, is set equal to
$\alpha P$.  The stresses observed in the simulations can be similarly
scaled using the vertically averaged pressure to derive an $\alpha$ as
a function of $R$.  In the simulations, $\alpha$ varies with both time
and space.  Its utility is mainly to provide a familiar point of
reference for characterizing the observed stress.  The local shearing
box simulations (HGB; HGB2) found that the Maxwell stress, which
dominates the total transport, is highly correlated with the total
magnetic pressure.  This suggests an alternate stress parameter, the
``magnetic alpha value'', defined as  $\alpha_m \equiv 2\langle B_R
B_\phi\rangle/ \langle B^2\rangle$.

In a steady state disk, the total stress, Maxwell plus Reynolds, would be
equal to radial angular momentum flux carried inward by the net radial
drift velocity, minus the angular momentum flux at the inner boundary
of the disk where the stress is presumed to vanish.  The latter term is
generally ignored for radii large compared to the inner disk boundary.
In the simulations none of these assumptions hold {\it a priori}:  the
typical torus radius is comparable to the inner disk boundary, the
stress does not vanish at the inner disk boundary, and the disk will
not be in steady state.
In fact, the simulations will prove to be highly dynamic
with rapid variations in both time and space.  This is best
appreciated from time-dependent animations which also have been used
as diagnostics for these global simulations.  While reporting time- and
space-averaged quantities tends to obscure this feature,
such averages are more practical for a  written summary.

\section{Results}

In this section the results from a range of three dimensional
simulations are presented.  The simulations are listed in the summary
Table which gives a model designation, the grid resolution used, the
gravitational potential, the initial torus angular momentum
distribution $q$, the initial magnetic field topology, and the time to
which the simulation was run.

\subsection{Cylindrical Disks}

The first set of simulations make use of a cylindrical gravitational
potential $\sim 1/R$ (the cylindrical disk limit).  This is a significant
simplification, as it permits the use of periodic boundary conditions in
the vertical direction.  Further, with this potential there is only one
important vertical lengthscale, namely that of the MRI; this reduces
the number of $z$ grid zones required.  Finally, in the cylindrical
limit one can study the evolution of models with vertical fields
without the stringent Courant limits due to the high Alfv\'en speeds
associated with strong fields passing through the low density region
above the disk.

The first two cylindrical simulations assume a Keplerian initial disk.
Keplerian simulations provide an immediate comparison with the
local shearing box results which also, for the most part, assumed
Keplerian flow.  Consider first the run labeled CK1 in the summary
Table; this is a cylindrical computational domain $(R,\phi,z)$ running
from 1 to 4 in $R$, 0 to 1 in $z$, and 0 to $\pi/2$ in $\phi$.  The
grid has $90\times 80\times 24$ zones.  An outflow boundary condition
is used at both the outer and inner radial boundaries, and periodic
boundary conditions are used for $\phi$ and $z$.  The initial disk has
a constant mass density from $r=1.5$ to the outer boundary.  The
adiabatic sound speed is constant and equal to 5\% of the orbital
velocity at the inner edge of the disk.  The initial magnetic field is
vertical and proportional to $\sin(R)/R$ between $R=1.5$ and 3.5, with
a maximum strength corresponding to $\beta=400$.  The strength of the
field, and hence the Alfv\'en speed $v_A$, was chosen so that the
characteristic wavelength of the MRI was $\lambda_{MRI} \equiv 2\pi
v_A/\Omega = 0.38$ at the location of the field strength maximum.  This
ensures that the $z$ domain size and the vertical grid resolution will
be adequate to resolve the fastest growing modes.  Random
nonaxisymmetric pressure fluctuations are added to create a full range
of low amplitude initial perturbations.

As the evolution proceeds, the MRI sets in, field energy is amplified,
and soon the characteristic radial streaming structures (referred to as
the channel solution) of the vertical field instability appear, much as
they do in the local shearing box models.  In the present simulation,
these structures develop first at the inner part of the disk where the
rotation frequency is the highest.  The amplitude of the MRI becomes
nonlinear by 3 orbits at the center of the grid $(P_{\rm orb} =
24.8$), and filaments of strong magnetic field are
carried inward and outward by fluid elements well out of Keplerian
balance.  These reach the outer part of the disk even before the local
MRI in that region becomes fully nonlinear.  Thus there are two
immediate global effects not seen in local simulations:  linear growth
rates that vary strongly with radius ($\omega_{MRI} \propto \Omega \propto
R^{-3/2}$), and extended radial motions of significantly non-Keplerian
fluid.

In local simulations with initial vertical fields that vary
sinusoidally with radius, the initial phase of the instability promptly
breaks down into MHD turbulence.  This happens in the global simulation
as well.  After the onset of turbulence, the disk displays many tightly
wrapped (i.e., large radial wavenumber) trailing spiral features with low
azimuthal wavenumber, $m$.  The magnetic field energy saturates at about
$\beta=4$, with the toroidal component dominant:  $B_R^2/B_\phi^2 =
0.075$ and $B_z^2/B_R^2 = 0.34$.  Similar ratios were found in the
shearing box simulations with zero net initial magnetic field (HGB2).

The MHD turbulence produces rapid angular momentum transport and mass
accretion, with $\alpha$ peaking at 0.21 at $t=90$, and declining
beyond this point, dropping to 0.06 at the end of the simulation.  The
Maxwell stress displays a high correlation with the total magnetic
pressure.  After an early peak at 0.68, $\alpha_m$ declines with time;
at the end of the simulation $\alpha_m = 0.35$.  The ratio Maxwell to
Reynolds stress is about 3 at the time of the largest total stress and
rises to about 9 at the end of the simulation.  The overall angular
momentum distribution remains nearly Keplerian throughout the
simulation, becoming slightly sub-Keplerian outside of the original
inner disk edge, and slightly super-Keplerian in the region between
this point and the inner grid boundary.

The mass flux, $\langle \dot M \rangle$, varies both in space and time.
After 8 orbits of evolution, over half of the initial disk mass has
been lost through the boundaries, particularly the outer boundary.
Since the initial Keplerian disk abuts the outer boundary, and the mass
increases with radius for a constant initial $\rho$, it is relatively easy for
a considerable amount of the mass to leave the grid,
driven in part by a pressure gradient created by a rarefaction wave.

This simulation is similar to that of Armitage (1998) except that it
uses a smaller domain in the angular direction, and a constant initial
density in the disk.  Qualitatively, the two runs exhibit similar
behavior, and both are consistent with results from local simulations.
One difference noted by Armitage, is that $\alpha \approx 0.1$ in the
global simulation; this is larger than the value of $\alpha \approx 0.01$
seen in local simulations with zero net vertical field.
The larger $\alpha$ is, however, consistent with what is seen in
local simulations with net vertical field.  The global simulations have
much larger radial and azimuthal wavelengths available to the MRI.  
Further, the radial scale
over which the vertical field sums to zero in the global simulation is
much larger than in the local shearing box.  
Here $B_z$ initially varies between $R=1.5$ and 3.5; this
range has a large variation in $\Omega$ and hence a large range in
growth rates of the MRI.  Finally, the presence of outflow radial
boundaries means the net field need not remain zero over the
simulation.  In the local simulation, periodic boundary ensure that the
net field cannot change, and by definition, the vertical field sums to
zero over a $\Delta R$ which is small.  Reconnection in the local box
should be much more efficient, reducing the total magnetic pressure and
hence the total stress.

The next simulation, labeled CK2, begins with the same Keplerian disk,
but with an initial toroidal field of constant strength, $\beta=4$,
lying between $r=2$ and 3.5.  Generally speaking, toroidal field
simulations are more challenging than vertical field simulations.
Although the presence of the linear instability is independent of the
orientation of the background magnetic field, the fastest growing
toroidal field modes are associated with large vertical and radial
wavenumbers.  Further, the critical azimuthal wavenumber is large,
unless the magnetic energy is comparable to the rotation energy.  Here
$m_{crit} \equiv 2\pi R/\lambda_{MRI} \approx 40$.  Since there are 80
azimuthal grid zones over $\pi/2$ in angle, the critical wavenumbers of
the instability will be resolved for this choice of field strength.
But it is clear that it takes a relatively strong toroidal field to do
so for a reasonable number of grid zones.  This illustrates one of the
difficulties with global models:  it is necessary to reconcile the
practical limitations of grid resolution with the length scales
associated with the weakest, and hence most interesting, magnetic
fields.

With a toroidal initial field, the disk evolves at a slower rate
that CK1.  The instability grows over the first few orbits,
with the fastest rate of growth associated with the innermost radius.
The growing modes of the instability have the same appearance as seen
in the local simulations; high $m$ structures appear first, building to
lower $m$ with time.   The vertical and radial structure also features
high wavenumbers.  The rapid growth phase of the instability ceases
after about 8 orbits at the grid center.  At this
point the field exhibits a low $m$, tightly wrapped structure, with
rapid variations in $R$ and $z$.  This behavior is consistent with the
linear analysis (Balbus \& Hawley 1992).  Turbulence and accretion
begin, and by 12 orbits over half of the disk mass has been lost from
the grid (much of it through the outer boundary); what remains is piled
up near $R=1.7$ and is slowly accreting.  The magnetic field is
subthermal and dominated by the toroidal component.  The component
magnetic energies at late time are quite similar to those seen in the
vertical field run CK1.

Angular momentum transport is again mainly due to the Maxwell stress.
Compared with CK1, the toroidal field model has lower overall stress.
This is partly because the vertical field model has a very vigorous
initial phase associated with the saturation of the linear
instability.  At the end of both CK1 and CK2, $\alpha_m =
0.3-0.4$.  The mass accretion rate, magnetic field strength, and
$\alpha$ fluctuate considerably both in time and in space.  The total
$\alpha$ value rises to 0.12 and then declines to 0.06 by the end of
CK2.  The average ratio of the Maxwell to Reynolds stress is 6 at late
times.  The value of $\alpha$ never rises as high as in local vertical
field models, including CK1.  It is comparable, however, to the typical
value found for weak field toroidal field models in the local shearing
box simulations that began with comparable initial toroidal field
strengths (HGB).

These Keplerian disk models suffer from significant mass loss through
the outer boundary.  In contrast, tori can be completely and
self-consistently contained initially within the computational domain.  We next
turn to cylindrical models of constant angular momentum ($q=2$) tori,
beginning with a torus with $R_{in} = 2.0$ and $R_{kep} = 2.5$.  The
torus outer boundary is at $R=3.3$.  Two simulations are done, model
CT1 with an initial vertical field, and CT2 with an initial toroidal
magnetic field;  both use the same computational grid as above.

Model CT1 begins with a vertical field with constant $\beta = 100$ from
$r=2.1$ to 3.1.  This gives $\lambda_{MRI} = 0.25$ at the pressure
maximum.  As the evolution proceeds, the magnetorotational instability
rapidly develops, again with the characteristic channel.  Early on, the
beginnings of the Papaloizou-Pringle principal mode can also be seen as
$k_z = 0$ oscillations at the edges of the torus.  But long before this
global instability can develop, the torus is dramatically altered by
the local MRI.  The torus does not endure as a torus: it expands
rapidly outward as the angular momentum distribution shifts from
constant toward Keplerian.  After 4 orbits at
the initial torus pressure maximum the system has evolved to a nearly
Keplerian disk that fills the computational domain.

The magnetic field is amplified to an overall average value of
$\beta=2$.  Toroidal field dominates:  $B_r^2/B_\phi^2 = 0.1$, and
$B_z^2/B_r^2 = 0.3$.  As always, the Maxwell stress exceeds the
Reynolds stress by a factor of several.  The value of $\alpha_m$ peaks
at 0.7 and declines to 0.4 at the end of the simulation ($t=102$).  The
overall $\alpha$ value varies throughout the disk; at $t=85$ it varies
between 0.3 and 0.4.

Next take the same torus and apply an initial toroidal field with
$\beta=100$ (Model CT2).  With this strength field, the critical
azimuthal wavenumber at the pressure maximum is $m_{crit}=63$, so the
fastest growing modes are underresolved on this grid.  As with the
Keplerian simulation CK2, the initial toroidal field model becomes
turbulent at a later time compared with an initial vertical field
model.  The total poloidal magnetic field amplification is also
considerably smaller than seen in the vertical field model.  Despite
this, the qualitative outcome of the instability for the torus is
largely the same.  The slower onset of the MRI allows the principal
mode of the Papaloizou-Pringle instability to appear, but soon the
transport of angular momentum brings this to a halt.  The disk spreads
outward with the bulk of the mass slowly moving inward.  The overall
angular momentum distribution changes from constant to Keplerian from
the inner boundary to $R=2.5$, and sub-Keplerian but increasing beyond
this point.  As in the previous simulations, the disk exhibits large
local fluctuations in density, stress and field strength, and many low
azimuthal, high radial wavenumber features.  Similar time and space
variations were observed in the local models as well; they are
characteristic of the MHD turbulence.

At late time the magnetic energy has risen to $\beta=15$, with the
toroidal field dominant by a large factor:  $B_r^2/B_\phi^2 = 0.04$.
The ratio of the vertical to radial field energy is 0.16.  The value of
$\alpha_m$ rises to $\sim 0.3$, close to, but below, the value in the
Keplerian toroidal field simulation above.  The total stress
corresponds to an $\alpha = 0.02$-0.03 at late time.  Again the Maxwell
stress exceeds the Reynolds component by a factor of 3 to 4.  These
properties are consistent with toroidal field shearing box simulations
(HGB).

Finally, consider a constant angular momentum torus, CT3, with an
initial field constructed by setting the azimuthal component of the
vector potential equal to the density in the torus, $A_\phi =
\rho(R)$.  The resulting field is normalized to an average $\beta$ of
100.  In the cylindrical limit, the density depends only on radius $R$
and the resulting field is vertical with zero net value integrated over
the disk.  This model also adopts the pseudo-Newtonian potential $\sim
1/(R-R_g)$ with $R_g=1$. The computation domain runs from $R=1.5$ to
13.5, $\phi=0$ to $\pi/2$ and $z=0$ to 2.  The grid resolution is
$128\times 64\times 32$.  The torus has an inner edge at $r=4.5$ and a
pressure maximum at $r_{kep} = 6.5$ where $P_{orb} = 88$; the orbital
period at the outer boundary is 289.  A new feature in this model is
the presence of the marginally stable orbit at $R_{ms}=3$.  Inside of
this radius, matter will rapidly accelerate inward.

The MRI grows rapidly in the inner regions of the disk, again with the
characteristic radial channel appearance.  Accretion through the inner
boundary begins at about $t=100$.  The magnetic energy rises to
peak at $\beta = 8$ at $t=150$.  The magnetic energy grows more slowly
after that point; additional small peaks are observed due to the growth
of the MRI in the outer parts of the disk.  Mass loss through the outer
boundary begins at $t=200$, after which the initial linear growth
phase is over and the disk is fully turbulent.  Between $t=200$ and the
end of the simulation, $\beta$ decreases from 10 to 6.  As always, the
toroidal field exceeds the radial field, with $B_r^2/B_\phi^2 = 0.16$.
The radial field is, in turn, greater than the vertical field,
$B_z^2/B_r^2 = 0.47$.

The overall Maxwell stress in the MHD turbulence reaches a peak at
$t=150$, drops off, and then slowly climbs again.  At late time
$\alpha_m = 0.5$, and the total stress parameter $\alpha$ varies with
radius inside the disk, from 0.1 near the initial pressure maximum, to
0.2 at the outer regions.  The angular momentum distribution rapidly
evolves away from constant, toward Keplerian.  By $t=420$ it is nearly
Keplerian from the pressure maximum inward, and sub-Keplerian out
beyond this point.  Inside of $R_{ms}=3$ the Maxwell and Reynolds
stresses decline with radius, although not as fast as the pressure.
This makes $\alpha$ rise sharply.

By the end of the simulation ($t=420$) over one quarter of the disk
mass has been lost, and 70\% of it has gone in past the marginally
stable orbit.  An examination of the various vertically- and
azimuthally-averaged velocities at the end of the run (Fig. 2) reveals
that the inflow velocity $v_r$ accelerates rapidly inside of $R=4$.  By
the time the inner grid boundary is encountered, the inflow speed
strongly supersonic and super-Alfv\'enic.  Inside the disk the radial
speed an order of magnitude smaller than the sound speed, and smaller
than the toroidal or poloidal Alfv\'en speeds, which themselves remain
subthermal.  Going inward from the marginally stable orbit the toroidal
Alfv\'en speed rises more slowly that the poloidal Alfv\'en speed.  The
radial field strength is roughly constant, but the fluid density drops
inside of $R_{ms}$ as the flow accelerates inward.  The specific
angular momentum drops 3\% from $R=3$ to the inner boundary at $R=1.5$
indicating that there is still some net stress inside $R_{ms}$.

\subsection{Axisymmetry:  Simulations in the (R,z) plane}

The cylindrical disk limit represents a useful way of simplifying the
full global problem.  Another potentially useful simplification is the
axisymmetric limit.  In this series of simulations, the torus evolution
problem is considered in the axisymmetric $(R,z)$ plane.  These
simulations use a pseudo-Newtonian potential, and begin with a
pressure-supported torus that is fully contained on the grid.  The
initial magnetic field is chosen to satisfy two requirements:  it must
have a poloidal component to allow for the development of the MRI, and,
as a practical matter, it should be contained completely within the
torus to avoid the Courant limitations caused by high Alfv\'en speeds
due to strong fields in low density regions.  A suitable initial
configuration consists of magnetic field loops lying along equidensity
surfaces in the torus.  This initial setup will develop strong toroidal
field due to shearing of the initial radial field.  However, experience
has shown that strong toroidal fields are the outcome of all initial
field choices, so this should not represent too great an idealization.
All of the two-dimensional simulations considered here will also be run
in three dimensions.

The first simulation has a radial grid running from $R=1.5$ to an outer
boundary at $R=11.5$ and a vertical grid running from $z=-5$ to 5.  The
grid resolution is $128\times 128$.  A constant-$\ell$ torus is placed
on the grid with an inner boundary at the marginally stable orbit,
$R_{in}=3$ and a pressure maximum at $R_{kep} = 4.7$.  The orbital
period at the pressure maximum is 50.  The initial field is obtained
from an azimuthal vector potential $A_\phi = \rho(R,z)$, and is
normalized to $\beta=100$ using the total integrated magnetic and
thermal energies.

Density grayscale plots from this simulation are presented in Figure
3.  The initial period of evolution is dominated by the growth of
toroidal magnetic field due to shear.  As the magnetic pressure
increases, the torus expands, particularly at the inner edge.  At
first, low density material is driven outward perpendicular to the
torus surface; subsequently, it flows radially out and around the
torus.  Because of the initial reflection symmetry across the equator,
the toroidal field changes sign there and a strong current sheet
forms.  This current sheet proves to be unstable, and oscillates around
the equator.  This is an important part of the evolution, and indicates
that it is necessary to simulate the full $(R,z)$ plane rather than
adopt the equator as an explicit boundary.

As the toroidal field pressure increases, it drives inflow through the
marginally stable orbit; this initial accretion flow peaks at $t=50$
then begins to decline.  In the meantime, the poloidal field MRI grows
within the torus, and begins to manifest itself visibly in the typical
form of  radial channels by $t=180$ (3.6 orbits).  There follows a
period of violent readjustment within the disk, featuring strong mass
inflow punctuated by episodic accretion events.  This phase lasts until
$t\sim 350$ (7 orbits), beyond which the poloidal magnetic energy
declines, and with it the level of turbulence in the disk, and the
accretion rate.  At the end of the simulation ($t=850$, $17$ orbits)
about 60\% of the initial disk mass has been lost.   Most of this mass
is lost by $t=500$; after this time the inflow accretion rate is very
small.

Thus there are three distinct phases to the two-dimensional torus
evolution:  expansion due to the shear-amplified toroidal magnetic
pressure, strong nonlinear evolution of the poloidal field MRI, and
finally a more quiescent turbulent state with declining poloidal
magnetic field strengths.  Angular momentum transport occurs in all
three phases at different rates.  In the first phase there is a growing
Maxwell stress from the amplified $B_\phi$ field, mainly in the inner
region of the disk where the orbital frequency is highest.  The
Reynolds stress is negligible.  The initially constant angular momentum
distribution is unchanged within the disk except in this inner region.
With the onset of the MRI, angular momentum transport occurs
everywhere, and the specific angular momentum begins to increase with
radius.  During the middle of this second phase, angular momentum
transport peaks, with $\alpha$ ranging between 0.2 and 0.5 through the
disk.  As the turbulence subsides, the rate of transport declines with
$\alpha$ varying between .02 and .1 within the main part of the disk.
Since this is an axisymmetric simulation, the strength of the poloidal
magnetic field is limited topologically by the antidynamo theorem, and
the MHD turbulence must eventually die out.

The next model is a torus located initially at a larger radius, which
increases the amount of time that it can evolve prior to reaching the
marginally stable orbit.  The grid runs from $R=1.5$ to $R=13.5$ and
from -4.5 to 4.5 in $z$.  The grid resolution is $120\times 90$.  The
constant-$\ell$ torus has an inner edge at $R_{in}=4.5$ and a pressure
maximum at $R_{kep} = 6.5$.  The orbital period at the pressure maximum
is 88.  Again, magnetic field loops are placed along the equidensity
surfaces within the torus, with a total magnetic energy corresponding
to $\beta = 100$.

As before, this torus undergoes three phases of evolution:  toroidal
field amplification, development of the nonlinear poloidal MRI, and
subsequent turbulence.  The MRI saturates around $t=300$ (3.4 orbits)
with total magnetic energy $\beta = 2$. The phase of strong MRI
turbulence is over by $t=450$ (5.1 orbits) when the average $\beta$
rises to 6; fairly steady accretion follows for the remainder of the
simulation which runs to time $t=830$ (9.5 orbits).

In the third simulation, the initial torus has an angular momentum
distribution closer to Keplerian, specifically $q=1.68$.  The torus
inner boundary is at 4 and its pressure maximum at $R_{kep}=10$
(orbital period $=179$).  The computational domain runs from
$R_{in}=1.5$ to 21.5, and from $-10$ to 10 in $z$; the grid resolution
is $128\times 128$.

Although it began with a much different initial angular momentum
distribution, the evolution is quite similar to the two previous
cases.  In the early stage, toroidal field is amplified by shear.  The
poloidal MRI soon comes into play, producing the characteristic radial
streams.  The total magnetic energy peaks at about $t=800$ (4.5
orbits).  After this the poloidal field energy declines steadily, as
does the Maxwell stress and the mass accretion through the inner radial
boundary.  Toroidal field dominates, with the ratio $B_r^2/B_\phi^2 =
0.017$.  Beyond $t=2000$, the toroidal field energy is essentially
constant with a $\beta= 1.3$.

Angular momentum transport begins almost immediately.  Initially it is
confined to the inner regions of the torus where shear is strongest,
and dominated by the Maxwell stress associated with the growing
toroidal field.  However, the growth of the MRI produces stress
throughout the torus.  By $t=1000$ the averaged $\alpha$ value is
between 0.1 and 0.2.  At the end of the simulation at $t=2840$,
$\alpha$ ranges between 0.01 and 0.04 within the torus.  The ratio of
the Maxwell stress to the magnetic pressure, $\alpha_m$, begins small,
rises to a value of 0.1 during the initial saturation of the MRI, then
steadily declines to 0.03 by the end of the simulation as poloidal
field is preferentially destroyed.  Since this torus began with an
angular momentum distribution similar to the end state of the initially
$q=2$ tori, there is only a slight change in the slope of the average
angular momentum.   Once the torus expands, this slope is extended in
range all the way from $R_{min}$ to the outer radial boundary.

\subsection{Three Dimensional Tori:  Full Global Simulations}

Now we turn to the evolution of fully three dimensional tori using a
pseudo-Newtonian potential.  The simulations described in this section
are three-dimensional versions of the tori considered above in the
axisymmetric limit.  The first, model GT1, is the constant angular
momentum torus with $r_{kep}=4.7$ and $r_{in}=3$.  The orbital period
at the pressure maximum is $t_{orb} = 50$.  The computational domain
runs from 1.5 to 11.5 in radius, from $-5$ to $5$ in $z$, and from
$\phi =0$ to $\pi/2$.  The initial magnetic field is constructed by
setting the toroidal component of the vector potential equal to the
density inside the disk, $A_\phi = \rho (R,z)$, for all $\rho$ greater
than a minimum value.  The total magnetic energy is then normalized to
a value of $\beta=100$, using the total integrated gas pressure of the
torus.  Different regions within the torus will, of course, have larger
or smaller values of $\beta$.  The strongest initial fields are found
between the pressure maximum and the inner edge of the torus.  The
simulation is run to time $t=780$ (15.6 orbits).

The evolution of this torus is illustrated with a series of 
grayscale plots in $\log (\rho)$ (Fig.~4) of vertical $(R,z)$ and
equatorial $(R,\phi)$ slices.  As with the
axisymmetric torus (Fig.~3), the initial phase of evolution is
controlled by the shear amplification of the toroidal field.  However,
in three-dimensions this toroidal field is itself unstable to the MRI.
The characteristic large-wavenumber structures associated with the
instability quickly appear in the inner regions of the disk (by $t=80$).
This creates turbulence, and momentarily increases the mass accretion
rate over that seen in the axisymmetric simulation (Fig.~5).  By
$t=200$ the disk contains tightly wrapped, low $m$ trailing spiral
waves.  The presence of this nonaxisymmetric structure affects the
development of the poloidal field instability, eliminating the
organized radial streaming flows seen in axisymmetry.  The flow is
more turbulent and the mass accretion rate is steadier,
without the large impulsive spikes in $\dot M$ associated with the
radial streaming in two dimensions.  This overall turbulence declines
after saturation of the MRI, but continues through the
end of the run.   This is the stage when the accretion rate in the
two-dimensional simulation drops to very small levels.  While in three
dimensions one might hope that the disk could achieve a nearly steady
state mass accretion rate, there is, of course, only a finite amount of
disk mass available to accrete.  At the end of the run 85\% of the
initial total mass has been lost from the grid, most of it through the
innermost stable orbit.  The amount of mass ejected off the outer
boundary is about 30\% of that accreted through the inner boundary.

Although the total volume-averaged magnetic energy drops with time
beyond $t=400$, most of this decline is due to loss from accretion.
Average $\beta$ after $t=200$ is $\approx 4$ and remains fairly
constant thereafter.  The magnetic energy is dominated by the toroidal
component; the poloidal field $\beta$ value has an average value of
$\sim 30$, and also remains nearly constant after $t=200$.  In the
axisymmetric simulation the poloidal $\beta$ value attains a value of
$15$ at $t=280$, but rises steadily to $\beta=75$ by $t=800$,
indicating a poloidal magnetic energy loss rate exceeding that due to
accretion alone.

Angular momentum transport results in rapid restructuring of the disk.
The angular momentum distribution in the inner half of the torus
steepens to a nearly Keplerian value by $t=50$ (Fig.~6).  The angular
momentum in the outer part of the torus increases over the remainder of
the simulation.  At the end time the angular velocity $\Omega \propto
R^{-q}$ has a slope near $q=1.55$.  This is essentially the same as in
the axisymmetric run above (represented by the short dashed line in
Fig.~6).  This slope adequately characterizes the disk since the angular
momentum, and hence angular velocity, is nearly constant on cylinders.
After turbulence has developed, the average $\alpha$ values
in the main part of the disk are between 0.1 and 0.2.  The ratio of the
Maxwell to Reynolds stress varies, but lies between 1 and 4 throughout
the main part of the disk.  The time- and space-averaged value of
$\alpha_m$ after $t=200$ is 0.40.

Figure 6 shows that the specific angular momentum continues to decline
even inside the marginally stable orbit.  This indicates that even here
there remains a significant net stress.  In fact, inside the marginally
stable orbit $\alpha$ rapidly increases because the gas pressure drops
while the Maxwell stress remains roughly constant.  The presence of
this stress means there is no maximum in the epicyclic frequency,
$\kappa^2 \equiv 2\Omega/r \ \partial \ell/\partial r$.  In the
pseudo-Newtonian potential (and, of course in the relativistic
potential that the pseudo-Newtonian potential was designed to imitate),
the Keplerian value of $\kappa^2$ has a maximum that occurs at about
$3.7 r_g$.  More generally, the epicyclic frequency will go to zero at
a stress-free inner edge of a disk where $\partial \ell/\partial r =
0$, ensuring a $\kappa$ maximum somewhere in the disk.  Here, however,
the inner boundary of the disk is not characterized by a zero-stress
condition, the slope of $\partial \ell/ \partial r$ is nearly constant,
and the epicyclic frequency does not turn over but continues to rise.
It has been proposed that potentially observable oscillatory modes
might be trapped near the disk inner edge where the epicyclic frequency
turns over (e.g., Nowak \& Wagoner 1993).  The present result must be
regarded as preliminary, but if significant stress inside the
marginally stable orbit proves to be a generic property of magnetic
turbulence, trapped disk oscillations may not be present.

In some respects, the two- and three-dimensional simulations are
similar.  They both have shear amplification of toroidal field, they
both evolve due to the resulting increase in toroidal magnetic
pressure, and they both develop turbulence due to the poloidal field
MRI.  Both rapidly evolve from constant to nearly Keplerian specific
angular momentum distributions (Fig.~6).  The three dimensional
simulation, however, permits the development of the nonaxisymmetric MRI
which increases and sustains turbulence and mass accretion.  In two
dimensions the organized poloidal field channel solution produces an
impulsive accretion rate greater than that seen in three dimensions
during the initial saturation of the poloidal MRI.  But axisymmetry
causes the two-dimensional poloidal field to decline, and with it the
Maxwell stress.  The contrast between $\alpha_m$ in the three- and the
two-dimensional simulations is instructive (Fig.~7).  In three
dimensions $\alpha_m$ remains relatively constant and near the value
typically seen in the local shearing box simulations.  In two
dimensions $\alpha_m$ declines steadily with time following the
saturation of the MRI.  Maintenance of the poloidal field through
dynamo action is possible only in three dimensions.

Although two-dimensional simulations cannot capture these essential
features of global evolution, they do have one clear advantage:  they
are considerably easier to compute.  Two dimensional simulations are
useful for searching a wide range of initial conditions in support of
the more challenging three-dimensional models.  To test the idea of
using an evolved two-dimensional simulation as an initial condition,
consider next a constant-$\ell$ torus with an initial inner edge at
$r_{in}=4.5$ and a pressure maximum at $r=6.5$ where $P_{orb} = 88$.
This same initial torus was run both in the cylindrical limit, and in
the axisymmetric $(R,z)$ limit above.  Here the computational domain
runs from $R=1.5$ to 13.5, $\phi=0$ to $\pi/2$ and $z=-4.5$ to 4.5.
The grid resolution is $120\times 64\times 90$.  The three dimensional
torus GT2 is initialized by taking the output from the axisymmetric
simulation at time $t=233$ and expanding it out in the $\phi$
direction.  This time corresponds to shortly before the nonlinear aspects
of the MRI first manifest themselves noticeably in the density plots.
Random pressure perturbations are added to break the axisymmetry.

While this initialization procedure reduces the total computational
time required, it doesn't allow the toroidal field instability the
opportunity to grow during the initial phase.  This means that the
strong coherent poloidal field instability develops as it does in
two dimensions.  Nevertheless, significant nonaxisymmetric structure
appears by $t=400$, and the overall turbulence is increased over that
seen in axisymmetry.  Accretion inflow at the inner boundary is about
twice what it is in two dimensions after this point in time.  After
$t=400$ the average magnetic field strength is $\beta=4.7$, and the
poloidal field strength $\beta_p = 24$.  At late time $\alpha \approx
0.1$, and $\alpha_m = 0.37$, a level similar to that seen in GT1.  In
two dimensions the Maxwell stress drops and the poloidal $\beta$
increase.  A the end of the simulation, $\alpha_m = 0.17$ and $\beta_p
= 55$.  The toroidal field $\beta$ is fairly constant with time in both
two- and three-dimensions runs after $t=450$.

This run can also be compared with the cylindrical run CT3.  Figure 8
is a plot of the radial dependence of the azimuthally and vertically
averaged speeds.  These curves are very similar to those in Figure 2
from CT3.  GT2 shows a stronger initial field amplification phase and a
stronger MRI saturation (created by initializing from the axisymmetric
run).  However the magnetic field amplitudes and stress levels are
comparable at late times in both simulations.  These similarities
indicate that the cylindrical limit provides a useful approximation
for investigating the nonlinear evolution of MHD turbulence near the
midplane of a disk.

To summarize, GT2 shows that fully three dimensional turbulence can be
rapidly produced and sustained in a simulation initialized from the
output of an axisymmetric simulation.  The late-time properties of such
a simulation, essentially one with complicated nonlinear initial
perturbations, are quite similar to those obtained from a simulation
evolved from a formal equilibrium and linear perturbations.
The two-dimensional simulations, therefore, serve as just another type
of initial condition, albeit more complicated than usually adopted.

One characteristic of all these simulations is that at late times the
magnetic field is predominantly toroidal.  The next simulation
considers  a torus that begins with only a toroidal field.  Model GT3
is of a $q=2$ torus with an inner boundary at 4.5 and a pressure
maximum at $R_{kep}=6.5$ (orbital period  $=88$).  The initial toroidal
field has $\beta=4$ everywhere within the torus.  The MRI that will
result is nonaxisymmetric; an axisymmetric $(R,z)$ version of this
model does not develop turbulence or transport angular momentum.  The
computational domain runs from $R_{in}=1.5$ to 13.5, $z$ from $-6$ to
6, and over the full $2\pi$ in angle; the grid resolution is $128\times
128\times 128$.

No attempt is made to keep the initial torus in pressure balance;
the initial magnetic field simply adds to the equilibrium hydrodynamic
pressure.  As a consequence, the torus undergoes an
axisymmetric readjustment due to this additional magnetic pressure.  The
toroidal field MRI develops rapidly at the inner edge of the disk
where $\Omega$ is largest, before spreading throughout the disk.  The
poloidal field energy grows steadily with time until about $t=250$ when
it reaches a value of $\beta_p = 34$.  After $t=400$, $\beta_p$ increases
with time to 65 at the end of the run.  Grayscale images of GT3 are given
in Figure 9.

Models GT2 and GT3 began with the same hydrodynamic torus; they differ
in their initial magnetic fields, in the size of the $\phi$ domain, and
in resolution.  Despite these differences, the late time evolution in 
these two runs is very similar.  
They have comparable mass accretion rates through
$R_{min}$ (Fig.~5).  Both have $\alpha_m = 0.36$ at $t=720$.  Figure
10 illustrates the redistribution of mass and angular momentum as a
function of time.  The slope of the angular momentum distribution at
late time is the same in GT3 as in both GT1 and GT2.

Figure 11 shows the radial run of density $\Sigma$, velocity $\langle
v_R\rangle$, and mass flux $\langle \dot M\rangle$ at time $t=720$.
Density is normalized by the maximum $\Sigma$ at $t=0$.  Within the
torus the radial drift velocity is small compared to the orbital
velocity; $\langle v_R\rangle$ reverses outside of $R=8$ beyond which
there is net outflow.  Inside $R_{ms}$ the density decreases rapidly as
the inflow accelerates inward.  The accretion rate increases slightly
from $R=7$ inward.  Figure 12 shows gas and magnetic pressures at this
same end time (normalized by the initial pressure maximum value), as
well as the Maxwell stress.  The magnetic energy remains subthermal
throughout, although it drops less rapidly than the gas pressure inside of
$R_{ms}$.  The ratio of the Maxwell stress to the gas pressure has a
minimum of 0.03 between $R=4$ and 5, rising sharply inside $R_{ms}$ and
more slowly toward the outer boundary.

The final global simulation, GT4, is the $q=1.68$ torus with an inner
boundary at 4 and a pressure maximum at $R_{kep}=10$.  This angular
momentum distribution yields a torus that extends over a large radial
distance without becoming too thick in the vertical direction.  The
computational domain runs from $R_{in}=1.5$ to 21.5, from $-10$ to 10
in $z$, and over the full $2\pi$ in angle; the grid resolution is
$128\times 128\times 128$.  The simulation is run to time $t=1280$.
This is longer than the other simulations, although it amounts to fewer
orbits at the pressure maximum:  $t=1280$ is 7.2 orbits at $R_{kep}$.
This evolution time is sufficient to observe all of the stages seen in
the other three-dimensional simulations, although not long enough for
the torus to have settled into a quasi-steady state.

Density plots from GT4 are presented in Figure 13.  At the beginning,
the toroidal field grows by shearing out the radial field, but as it
does so, the toroidal MRI sets in.  This soon leads to turbulence.  The
poloidal MRI develops rapidly in the inner regions of the disk and
subsequently spreads throughout.  The inner edge of the disk moves
slowly inward until it reaches the marginally stable orbit at $t=145$
after which gas plunges inward.  Initially the inflowing fluid is confined 
largely to the equatorial plane.  As time passes, however, this region fills
with gas and becomes thicker.

As with the previous global simulations, the mass accretion rate is
larger in three dimensions compared to two.  Near the end of the
simulation, $\langle \dot M \rangle$ at the inner radial boundary is
approximately 2.5 times that seen in axisymmetry.  The value of
$\alpha_m$ is rising with time indicating that the poloidal field
strength is still increasing at the end of the simulation.

Figure 14 shows the radial mass and angular momentum distributions at
the initial and final times in GT4.  Also shown are curves from two
times in the equivalent axisymmetric calculation.  The average angular
velocity parameter $q$ decreases by a very small amount over the course
of the evolution.  The torus mass, on the other hand, has been substantially
redistributed.

\section{Discussion}

In this paper, we have carried out global MHD simulations of accretion
tori.  We have also made use of several different limits or
approximations:  two dimensional axisymmetry, and three dimensional
cylindrical gravitational potential,  in addition to the full three
dimensional global model.  In addition to the intrinsic interest in the
results, these simulations begin to map out what is currently possible
with the present hardware and algorithms.

First, how useful are the two-dimensional and cylindrical
approximations?  While neither should be relied on exclusively, they
both have appropriate applications, and they greatly reduce the
computational difficulty of the simulations.  Two-dimensional
axisymmetric torus simulations demonstrate effects from the generation
of toroidal field due to shear, and from the development of the
poloidal field MRI.  The latter leads to turbulence, rapid angular
momentum transport and evolution toward a nearly Keplerian angular
momentum distribution.  A significant limitation of the axisymmetric
system, however, is embodied in Cowling's antidynamo theorem.  The
$A_\phi$ component of the vector potential is conserved, except for
losses due to dissipative processes.  Poloidal field can grow from
axisymmetric stretching and folding, but this is ultimately limited.
In the simulations, after the nonlinear saturation of the MRI, the
poloidal magnetic field energies decline and the turbulence begins to
die out.

Despite this, considerable similarity is seen between the two- and
three-dimensional simulations.  This is due to the dominance in the
initial stages of the torus evolution by what is largely axisymmetric
dynamics:  increase in toroidal field pressure due to shear, the
development of the ``channel solution,'' and the rapid redistribution
of angular momentum by large stresses.  During phases of a disk's
evolution when such effects are most important, two-dimensional
simulations are a good approximation.  Their utility must be
limited, however, since genuine nonaxisymmetric effects, including the
development of the toroidal field MRI, dynamo amplification and
maintenance of poloidal fields, and nonaxisymmetric spiral waves, are
generally of dominant importance over the long term.

Cylindrical disks are a natural extension of the shearing box model.
The cylindrical disk, like the shearing box simulations of HGB and
HGB2, does not include the effects of vertical gravity.
Cylindrical Keplerian disk simulations, initialized with vertical
fields or toroidal fields show rapid development of the MRI consistent
with the shearing box results.  Field amplification and stress levels
are comparable between the two types of simulation as well.  
With the cylindrical disk, however, we are able to observe
the direct consequences of the stress:  redistribution of angular
momentum toward Keplerian, and mass accretion.  Cylindrical disk
simulations can provide detailed information about the radial
dependence of the growth and saturation of the MRI, the resulting MHD
turbulence, the transport of angular momentum, and the net accretion
flow.

Simulations of cylindrical constant angular momentum tori with vertical
or toroidal initial fields illustrate many aspects of a full global
evolution.  Regardless of the initial field configuration, the
initially constant angular momentum tori evolve toward radial angular
momentum distributions that are nearly, but not quite, Keplerian.  The
final outcome is largely the same for either initial vertical or
toroidal fields, although the early growth of the MRI is more rapid for
vertical fields, and dominated by small scale structure for toroidal
fields.  This is entirely consistent with the local linear analysis.
Of course, we want to simulate disks fully globally and with
as few approximations or geometric constraints as possible.  Indeed,
full three dimensionality is essential.  Cylindrical simulations can provide
little, if any, information about the vertical structure of a disk, energy
transport, or the possible formation of winds or jets.   

As with the cylindrical disk simulations, the overall evolution of
fully global tori are consistent with the intuitions developed from shearing
box simulations.  The MRI grows rapidly, and produces MHD turbulence
that transports angular momentum.  In all cases, toroidal field
dominates, followed by radial and then vertical field.  One difference
is that the shearing box calculations with zero net field (e.g., HGB2)
typically have a total stress value of $\alpha \approx 0.01$, whereas
here $\alpha \approx 0.1$ in the heart of the turbulent disks.  This is
more a matter of the magnetic pressure that is sustained in the torus
versus the shearing box, rather than some qualitative difference in the
behavior of the MRI.  In both the global and local simulations,  the
stress is directly proportional to $B^2/8\pi$:  $\alpha_m \approx
0.4-0.5$.  All of the present global simulations began with relatively
strong field, either in the form of poloidal field loops which
immediately generated strong toroidal field through shear, or from the
presence of an initially strong toroidal field.  This field strength is
by and large sustained, and thus the observed $\alpha$ values are
consistent with $\beta \le 10$.  It is suggestive that Matsumoto (1999)
also obtained $\beta \approx 10$ for his toroidal field simulations,
even one that began with $\beta=1000$.   Global (rather than local)
length scales make it natural to have larger magnetic field strengths
in the saturated state.

Computational problems associated with the inner boundary are greatly
reduced through the use of the pseudo-Newtonian potential.  This
provides a physical inner radius for the disks, and ensures that gas
and field will flow smoothly off the inner radial grid.  The accretion flow 
rapidly accelerates inward near $r_{ms}$, with the radial speed
quickly exceeding the sound speed.  Interestingly, the angular momentum
continues to drop inside of $r_{ms}$, indicating the continuing 
presence of Maxwell stress.  The absence of a stress-free inner disk 
boundary is one of the ways in which these disks differ from standard
analytic models.

One immediate conclusion from these simulations is that the constant,
or nearly-constant angular momentum torus is a remarkably unstable
structure in the presence of a weak magnetic field.  The MHD turbulence
resulting from the MRI simply transports angular momentum too
efficiently.  Within a few orbits the angular momentum distribution
changes from $q=2$ to $q\approx 1.6$.  The tori considered here did not
remain maintain their initial constant angular momentum distribution
long enough to develop the coherent structures
associated with the nonlinear outcome of the Papaloizou-Pringle
instability.  This does not, however, necessarily imply that accretion
disks must be thin.  Moderately thick configurations, as seen in the
grayscale plots, are still possible even for nearly Keplerian angular
momentum distributions, if there is significant internal pressure.

Averaged properties of the torus such as density, pressure or angular
momentum distribution fail to convey the impression of disorder seen in
animations of the evolution.  Low density, high magnetic field
filaments entwine themselves throughout the torus.  Regions of strong
field develop and rise through the torus into the surrounding low-density
atmosphere.  Low density material flows outward around the bulk of the
torus.  Since angular momentum transport is by MHD turbulence, it
follows that the disk should be highly dynamic.  The effect of this on
observed properties of disks must be considered in subsequent, more
sophisticated simulations.  It is not premature, however, to question
the relevance of the traditional image of a quiescent steady-state
disk.

There remain several limitations to the present simulations to be
addressed in subsequent work.  First, greater grid resolution is always
welcome, particularly where the accretion inflow is squeezed into a
narrow equatorial flow, or for following the growth of the MRI from
substantially weaker initial field strengths.  Because the disks
develop oscillations across the equatorial plane, particularly in the
simulations that begin with poloidal field loops, the full $(R,z)$
plane must be simulated in global models, without applying explicit
(e.g., reflecting) equatorial boundary conditions.  This doubles the
number of grid zones required, but appears to be necessary.  Higher
resolution simulations are feasible with the addition of more
processors to the parallel computation.  The simulations described here
used a maximum of 64.  Indeed, a great many issues can be profitably
investigated using simulations with more grid zones.  For example, the
global simulations presented here featured tori that began close to the
marginally stable orbit.  Little evolution is required to accrete
through $r_{ms}$ and into the central hole.  Simulating a greater
dynamic range in disk radii, for longer times and with larger numbers
of radial grid zones, is an immediate next goal.

The present simulations focus only on the dynamical properties of MHD
in tori.  As the thermodynamic properties of the disk are of obvious
importance in establishing an overall global disk structure, further
algorithm development is desirable.  In the present simulations the
simplified equation of state and lack of explicit resistivity mean that
the only sources of gas heating are artificial shock viscosity and
adiabatic compression.  Some amount of energy is necessarily lost
through the numerical reconnection of magnetic field.  Further, neither
radiative transport nor simple radiative losses were included.  These
and other issues of global disk evolution must be deferred to
subsequent work.

This work is supported in part by NASA grants NAG5-3058 and NAG5-7500.
Three dimensional simulations were run on the Cray T90 and T3E systems
operated by the San Diego Supercomputer Center with resources provided
through a National Resource Allocation grant from the National Science
Foundation.  The three-dimensional torus simulation GT4 was run by Greg
Lindahl on a cluster of 64 alpha processors, part of the Centurion
system developed by the Legion Project at the University of Virginia.
A portion of this work was completed while attending the program on
Black Hole Astrophysics at the Institute for Theoretical Physics,
supported in part by the NSF under Grant No. PHY94-07194.  I thank
Steve Balbus, Charles Gammie, James Stone, and Wayne Winters, and the
members of the ITP Black Hole Astrophysics program for helpful comments
and discussions.

\clearpage
\begin{center}
{\bf References}
\end{center}

\refindent Armitage, P. 1998, ApJ, 501, L189
\refindent Balbus, S.~A., \& Hawley, J.~F. 1991, ApJ, 376, 214
\refindent Balbus, S.~A., \& Hawley, J.~F. 1992, ApJ, 400, 610
\refindent Balbus, S.~A., \& Hawley, J.~F. 1998, Rev Mod Phys, 70, 1
\refindent Blaes, O.~M., \& Hawley, J.~F. 1988, ApJ, 326, 277
\refindent Brandenburg, A.,  Nordlund, \AA, Stein, R.~F., \& Torkelsson,
U. 1995, ApJ, 446, 741
\refindent Curry, C., \& Pudritz, R. 1996, MNRAS, 281, 119
\refindent Evans, C. R. 1986, in Dynamical Spacetimes and Numerical
Relativity, ed. J. M. Centrella (New York: Cambridge Univ. Press), 3
\refindent Evans, C.~R., \& Hawley, J.~F. 1988, ApJ, 332, 659
\refindent Goldreich, P., Goodman, J., \& Narayan, R. 1986, MNRAS, 221, 339
\refindent Goodman, J., Narayan, R., \& Goldreich, P. 1987, MNRAS, 225, 695
\refindent Goodman, J., \& Xu, G. 1994, ApJ, 432, 213
\refindent Hawley, J.~F. 1987, MNRAS, 225, 677
\refindent Hawley, J.~F. 1991, ApJ, 381, 496
\refindent Hawley, J.~F. \& Balbus, S.~A. 1991, ApJ, 376, 223
\refindent Hawley, J.~F. \& Balbus, S.~A. 1992, ApJ, 400, 595
\refindent Hawley, J.~F. \& Balbus, S.~A. 1999, in Astrophysical Discs,
eds. Sellwood, J. A. \& Goodman, J. (Cambridge Univ. Press:  Cambridge), 108
\refindent Hawley, J.~F., Gammie, C.~F., Balbus, S.~A. 1995, ApJ, 440,
742 (HGB)
\refindent Hawley, J.~F., Gammie, C.~F., Balbus, S.~A. 1996, ApJ, 464,
690 (HGB2)
\refindent Hawley, J.~F., \& Stone, J.~M. 1995, Comp Phys Comm, 89, 127
\refindent Matsumoto, R. 1999, in Numerical Astrophysics, eds. Miyama,
S., Tomisaka, K. \& Hanawa, T (Kluwer:  Dordrecht), 195
\refindent Matsumoto, R., \& Shibata, K. 1997, in Accretion Phenomena
and Related Outflows, eds. D. Wickramsinghe, L. Ferrario, \& G.
Bicknell (San Francisco: PASP)
\refindent Matsumoto, R., \& Tajima, T. 1995, ApJ, 445, 767
\refindent Norman, M.~L., Wilson, J.~R., \& Barton, R.~T. 1980, ApJ, 239, 968
\refindent Nowak, M.~A., \& Wagoner, R.~V. 1993, ApJ, 418, 187
\refindent Ogilvie, C., \& Pringle, J.~E. 1996, MNRAS 279, 152
\refindent Paczy\'nsky, B., \& Wiita, P.~J. 1980, A\&A, 88, 23
\refindent Papaloizou, J.~C.~B. \& Pringle, J.~E. 1984, MNRAS, 208, 721
\refindent Shakura, N.~I., \& Sunyaev, R.~A. 1973, A\&A, 24, 337
\refindent Stone, J.~M., Hawley, J.~F., Gammie, C.~F., \& Balbus, S.~A. 
1996, ApJ, 463, 656
\refindent Stone, J.~M., \& Norman, M.~L. 1992a, ApJS, 80, 753
\refindent Stone, J.~M., \& Norman, M.~L. 1992b, ApJS, 80, 791

\clearpage

\begin{figure}
\caption{A schematic view of the cylindrical coordinate
computational domain.  The radial $R$ dimension begins at a minimum radius
$R_{min}$, the vertical $z$ dimension is centered on the equator, and
the angular direction $\phi$ runs from 0 to $2\pi/m$ where $m$ is an
integer.  The simulations in this paper use $m=1$ and $m=4$.
}
\end{figure}

\begin{figure}
\caption{Vertically- and azimuthally-averaged velocities 
as a function of radius in simulation CT3 at time $t=420$ 
(4.8 orbits at the initial pressure maximum).
The curves trace the toroidal speed $v_\phi$, the adiabatic
sound speed $c_s$, the toroidal and poloidal Alfv\'en speeds $v_{A\phi}$
and $v_{Ap}$, and the radial speed $v_r$.
The vertical dashed line indicates the location of the marginally
stable orbit in the pseudo-Newtonian potential.  The short-dashed
line corresponds to the Keplerian velocity.
}
\end{figure}

\begin{figure}
\caption{Grayscale plots of $\log(\rho)$ in a two-dimensional
axisymmetric simulation of a constant angular momentum torus
containing poloidal field loops.  Each image is labeled by time.  At
$t=50$ the torus has expanded due to shear amplification of the
toroidal field.  At $t=180$ the poloidal field MRI has set in.  A
period of turbulence follows ($t=250$, 350) which dies out by the
end of the simulation ($t=850$, 17 orbits at the initial pressure
maxiumum).
}
\end{figure}

\begin{figure}
\caption{Grayscale plots of $\log(\rho)$ in simulation GT1, initially
a constant angular momentum torus
containing poloidal field loops.  Each image consists of a side view
and an equatorial view, and is labeled by time.  At
$t=80$ the torus has expanded due to shear amplification of the
toroidal field which has become visibly unstable in the inner regions.  Full
turbulence sets in by $t=200$ (4 orbits at the initial pressure maximum) 
and continues for the remainder of the
simulation.  The total disk mass drops steadily due to accretion.
These images should be compared with the axisymmetric model in Figure 3.
}
\end{figure}

\begin{figure}
\caption{Mass flux through the inner grid radius $R_{min}$ as a function of 
time for global models (solid lines) and their axisymmetric counterparts
where applicable (dashed lines).  In each case, the mass accretion rate 
is normalized
by the initial mass of the torus.  In models GT1, GT2 and GT4, the
initial accretion is driven mainly by the growth of the toroidal field
due to shear.  Subsequently the poloidal MRI drives strong accretion.
The coherent channel solution in axisymmetry produces particularly
strong accretion events.  At late times, accretion in the axisymmetric
models dies down while in the global models accretion continues at a
reduced, but significant, level.  Note that
despite their quite different initial field strengths and topologies, 
models GT2 and GT3 have similar accretion rates at late times.
}
\end{figure}

\begin{figure}
\caption{Mass distribution as a function of radius (top), and
averaged angular momentum as a function of radius 
$\langle \ell\rangle$ (bottom) at several different times in
model GT1.  The angular momentum curves are labeled by time; the corresponding
mass curves run from top to bottom with increasing time.  
The orbital time at the initial pressure maximum is 50.  The long dashed line
corresponds to the Keplerian value in the pseudo-Newtonian potential.
The short dashed line is $\langle \ell\rangle$ at time
$t=420$ in the axisymmetric version of this torus. 
}
\end{figure}

\begin{figure}
\caption{Time evolution of volume-integrated value of $\alpha_m\equiv
-B_rB_\phi/4\pi P_{mag}$ in the global torus models (solid lines) and
the axisymmetric version (dashed lines) where appropriate.  Individual
graphs are labeled by global torus model number.  In three dimensions,
poloidal field amplitudes are maintained relative to the toroidal
field, and the Maxwell stress remains appreciable compared to the total
magnetic pressure.  In two dimensions the poloidal fields die out and
the stress drops.
}
\end{figure}

\begin{figure}
\caption{Vertically- and azimuthally-averaged velocities 
as a function of radius at time $t=727$ (8.3 orbits at the initial pressure
maximum) in simulation GT2.
The curves trace the toroidal speed $v_\phi$, the adiabatic
sound speed $c_s$, the toroidal and poloidal Alfv\'en speeds $v_{A\phi}$
and $v_{Ap}$, and the radial speed $v_r$.
The vertical dashed line indicates the location of the marginally
stable orbit in the pseudo-Newtonian potential.  The short-dashed
line corresponds to the Keplerian velocity.
}
\end{figure}

\begin{figure}
\caption{Grayscale plots of $\log(\rho)$ in simulation GT3,
a torus with $q=2$ and $\beta=4$ toroidal field initially.
Each frame consists of a side
and an equatorial view, and is labeled by time.  At
$t=120$ the MRI has appeared at the inner
edge of the torus.  Turbulence is soon established throughout the
torus.
}
\end{figure}

\begin{figure}
\caption{Time evolution of the vertically- and azimuthally-averaged mass
(top) and specific angular momentum (bottom) as a function of radius in
simulation GT3.
The different lines correspond to  different times throughout
the evolution:  $t=0$, 117, 209, 298, 382, 470, 550, 637, 727.  The
orbital period at the initial torus pressure maximum is 88.  The long
dashed line in the angular moment plot corresponds to the Keplerian
value.  The short dashed line is the curve from the final time in model GT1.
}
\end{figure}

\begin{figure}
\caption{From top to bottom, the vertically averaged density, 
radial velocity, and mass flux as
a function of radius in model GT3 at $t=720$ (8.2 orbits at the initial
pressure maximum).  The dashed line
in the radial velocity plot is the orbital velocity.
}
\end{figure}

\begin{figure}
\caption{Vertically averaged gas pressure $P$, toroidal magnetic field
pressure $B_\phi^2/8\pi$, radial field pressure $B_R^2/8\pi$, and 
Maxwell stress as a function of radius at $t=720$ in model GT3.
The pressures are normalized to the initial pressure maximum value.
}
\end{figure}

\begin{figure}
\caption{Grayscale plots of $\log(\rho)$ in simulation GT4,
initially a torus with $q=1.68$ and
poloidal field loops.  Each frame consists of a side 
and an equatorial view, and is labeled by time.  The orbital period
at the initial pressure maximum is 179.  At
$t=320$ the torus has expanded due to shear amplification of the
toroidal field which has, in turn, become unstable with $m=4$ dominating
in the equatorial plane.
Beyond this time the poloidal field MRI begins and turbulence follows.
}
\end{figure}

\begin{figure}
\caption{Vertically and azimuthally averaged mass
(top) and specific angular momentum (bottom) as a function of radius 
for the initial and final time, $t=1283$ in model GT4.  The slope of 
$\langle \ell \rangle$ has steepened very slightly over the course
of the evolution.  The long
dashed line in the angular momentum plot corresponds to the Keplerian
value.  The short dashed lines are from times $t=1559$ and $t=2840$ in the
axisymmetric simulation of this torus. 
}
\end{figure}

\clearpage
\begin{center}
{\bf TABLE} \\ THREE DIMENSIONAL SIMULATIONS \\

\addvspace {0.5cm}

\begin{tabular}{cccccc} \hline \hline
\\ [-0.3cm]
Model&
Grid &
Potential &
q &
Initial Field  &
End time \\
\hline
\\ [-0.2cm]

CK1& $90\times 80\times 24$ & $1/R$ &1.5& $B_o \sin (R)/R \hat z$& 188  \\
CK2&$90\times 80\times 24$& $1/R$ &1.5&$B_o \hat \phi$& 285 \\
CT1&$90\times 80\times 24$& $1/R$ &2&$B_o \hat z$& 102 \\
CT2&$90\times 80\times 24$& $1/R$ &2&$B_o \hat \phi$& 403 \\
CT3&$128\times 64\times 32$& $1/(R-R_g)$ &2& $A_\phi \propto \rho(R)$& 420 \\
GT1&$128\times 64\times 128$& $1/(r-r_g)$ &2& $A_\phi \propto \rho(R,z)$& 780 \\
GT2&$120\times 64\times 90$&$1/(r-r_g)$ &2& $A_\phi \propto \rho(R,z)$& 727 \\
GT3&$128\times 128\times 128$&$1/(r-r_g)$&2&$B_o \hat\phi$& 727 \\
GT4&$128\times 128\times 128$&$1/(r-r_g)$&1.68&$A_\phi\propto\rho(R,z)$& 1283 \\
\hline
\end{tabular}
\end{center}

\end{document}